\documentclass[]{vgtc}                          

\graphicspath{{figures/}{pictures/}{images/}{figs/}{./}} 

\usepackage{times}                     

\usepackage{mathptmx}                  

\usepackage[dvipsnames,svgnames]{xcolor}
\usepackage{tikz}

\usepackage[nocompress]{cite}
\usepackage{array}
\usepackage{multirow}
\usepackage{ifthen}
\usepackage{subcaption}
\usepackage{siunitx}
\usepackage{amsmath}
\usepackage{enumitem}
\usepackage{ifthen} 
\usepackage{graphicx}
\usepackage{enumitem} 

\definecolor{bgMental}{HTML}{FFE6E6}   
\definecolor{bgTime}{HTML}{FFFBD1}     
\definecolor{bgWorld}{HTML}{E8F4FF}    
\definecolor{bgSpace}{HTML}{E8FFE8}    
\definecolor{bgSocial}{HTML}{F3E8FF}   
\definecolor{bgSelf}{HTML}{FFF1E0}     
\definecolor{bgResume}{HTML}{ECF0FF}   
\definecolor{bgInter}{HTML}{E0FFFF}    
\definecolor{bgCollab}{HTML}{FCE6F5}   

\definecolor{bgCtx}{HTML}{b3e2cd}
\definecolor{bgTask}{HTML}{cbd5e8}
\definecolor{bgTransition}{HTML}{fdcdac}

\newboolean{showrevisions}
\setboolean{showrevisions}{false}
\newboolean{isdraft} 
\setboolean{isdraft}{true} 

\newcommand{\secref}[1]{\hyperref[#1]{Section~\ref*{#1}}}
\newcommand{\figref}[1]{\hyperref[#1]{Figure~\ref*{#1}}}

\ifthenelse{\boolean{showrevisions}}{\definecolor{newcolor}{rgb}{0.15, 0.15, 1}
\definecolor{chcolor}{rgb}{0.15, 0.5, 0.15}}{\definecolor{newcolor}{rgb}{0, 0, 0}
\definecolor{chcolor}{rgb}{0, 0, 0}}

\ifthenelse{\boolean{isdraft}}{\newcommand{\todo}[2]{~\\ \textcolor{red}{\textbf{[\textsc{\textcolor{green}{#1}}: #2]}}}}{\newcommand{\todo}[2]{}} 
\ifthenelse{\boolean{isdraft}}{\newcommand{\note}[1]{\textcolor{red}{\textbf{[#1]}}}}{\newcommand{\note}[1]{}} 

\onlineid{0}

\vgtccategory{Research}

\vgtcinsertpkg

\author{Matt Gottsacker\thanks{e-mail: \{mattg, yohan.hmaiti, mykola.maslych, bruder, jlaviola, welch\}@ucf.edu}
\and Yahya Hmaiti$^*$%
\and Mykola Maslych$^*$%
\and Gerd Bruder$^*$%
\and Joseph J. LaViola Jr.$^*$%
\vspace{1.5mm}
\and Gregory F. Welch$^*$}%
\affiliation{\vspace{-2mm}SREAL and ISUE Lab, University of Central Florida \vspace{-5mm}} %

\title{
XR-First Design for Productivity: \\A Conceptual Framework for Enabling Efficient Task Switching in XR
}

\abstract{
A core component of completing tasks efficiently in computer-supported knowledge work is the ability for users to rapidly switch their focus (and interaction) across different applications using various shortcuts and gestures.
This feature set has been explored in research, and several modern consumer extended reality (XR) headsets now support loading multiple applications windows at once.
However, many XR applications that are useful for knowledge work involve rich spatial information, which window-based metaphors do not sufficiently represent nor afford appropriate interaction.
In modern XR headsets, such immersive applications run as siloed experiences, requiring the user to fully exit one before starting another.
We present a vision for achieving an XR-first, user-centric paradigm for efficient context switching in XR to encourage and guide future research and development of XR context- and task-switching interfaces.
}


\begin{document}


\firstsection{Introduction}
\maketitle

The promise of extended reality (XR) as a productivity platform, one that truly leverages the unique affordances of immersive spatial computing rather than replicating traditional window-based paradigms, hinges on users' ability to seamlessly transition between multiple tasks and contexts. This capability remains fundamentally lacking in current systems. While desktop users leverage millisecond-fast task switching through established window-management concepts, like Alt+Tab and creating 
working sets of applications, 
XR users face a jarring ``one-world-at-a-time'' paradigm that requires completely exiting an application before entering a new one.
This is more than a minor inconvenience.
Indeed, Gonzalez and Colaco's recent guidelines for XR productivity~\cite{gonzalez2024guidelines} highlight the lack of effective multitasking as one of several fundamental barriers to widespread adoption of XR for knowledge work.


Knowledge workers often need to transition fluidly between different tasks, from collaborative meetings to individual analysis, from information retrieval to 3D modeling, etc. 
In fact, knowledge workers switch between eight or more application windows~\cite{hutchings2004display} potentially hundreds of times every hour~\cite{olivier2006swish, leijten2014writing} to switch between and complete their tasks~\cite{jahanlou2023task, aral2012information}.
Recent commercial XR systems made progress in supporting multiple 2D applications simultaneously, but immersive applications remain siloed experiences~\cite{nam2019wedges}.
In addition to the documented cognitive overhead in XR productivity~\cite{biener2022quantifying}, we argue this ``one-world-at-a-time'' paradigm poses a significant barrier to broader XR adoption for daily knowledge work.

The root of this problem lies in the conservative application of desktop metaphors to immersive environments. Desktop window management was designed for flat, independent 2D applications that share consistent interaction paradigms (mouse + keyboard). XR applications, however, are fundamentally different in nature, necessitating an XR-first approach to design. XR applications involve spatial contexts, embodied interactions, and social presence
that cannot be adequately represented through traditional windowing systems. When users transition between immersive contexts, they must navigate changes in spatial orientation, embodiment, interaction modalities, and social awareness. We argue desktop metaphors fall short in addressing these complexities.



While prior research developed prototypes for XR context switching in controlled study settings, the field lacks a systematic understanding of how knowledge workers should manage tasks in immersive environments. We currently have no comprehensive framework that addresses the unique challenges posed by spatial context, embodied interaction, and social presence during task transitions. More critically, there is no systematic evaluation of design principles for task switching that could enable XR-first productivity with desktop-class performance.
In this paper we present an introductory conceptual framework for understanding and designing world management interfaces in XR knowledge work contexts. We analyze four representative knowledge work scenarios that reveal fundamental mismatches between current XR systems and user needs. We also present a preliminary workflow model that describes the factors affecting task switching in immersive environments, and discuss design challenges and opportunities for XR-first productivity solutions. 

In sum, we advocate for an XR-first approach to task switching that embraces spatial interaction, embodied collaboration, and environmental context as first-class design considerations rather than afterthoughts. Our goal is not to replicate desktop productivity features in XR, but to lay the groundwork for interaction paradigms that leverage the unique affordances of XR.
We hope that this work provides a starting foundation for future research and discussions that will help make XR a compelling platform for sustained knowledge work and day-to-day work.
\newcommand{\tightcolorboxfig}[2]{%
  {\setlength{\fboxsep}{1.25pt}\colorbox{#1}{#2}}%
}
\begin{figure*}[h]{
  \centering
  \includegraphics[width=\linewidth]{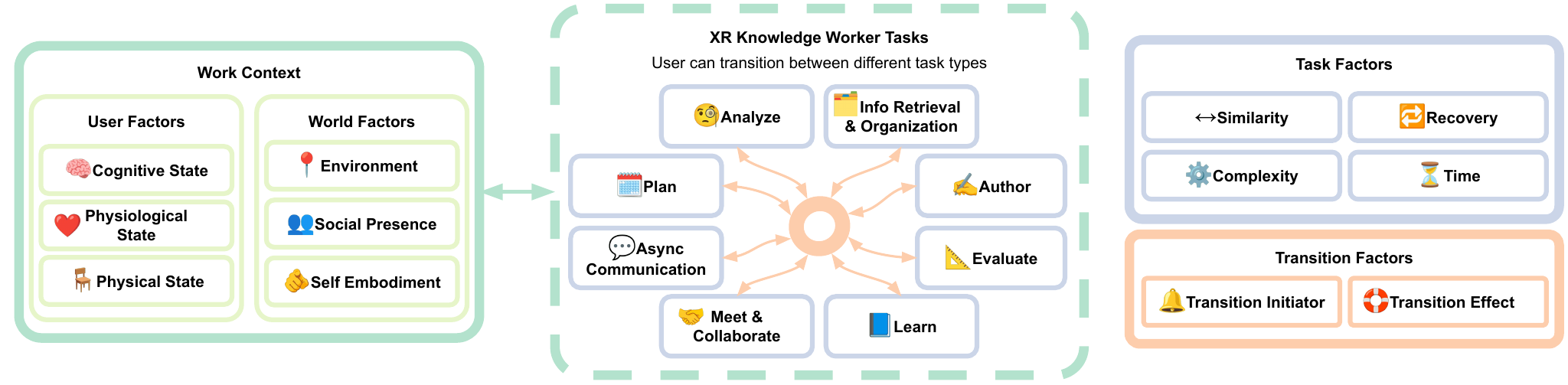}
  \caption{
  Preliminary conceptual framework for task switching in XR-first knowledge work.
  This framework highlights eight common knowledge work \tightcolorboxfig{bgTask}{tasks}, \tightcolorboxfig{bgTransition}{transitions} between them, and \tightcolorboxfig{bgCtx}{contextual factors} that shape user experience during task switching in XR. Every task switch engages task and transition factors and, in turn, dynamically shapes and is shaped by the overall XR context.}
  \vspace{-5mm}
  \label{fig:teaser}
}
\end{figure*}

\section{Task Switching in Knowledge Work}
Shortly after the advent of WIMP-based~\cite{vandam1997WIMp} GUIs on desktop computers, Bannon et al. identified an important problem: switching between concurrent tasks imposed so much cognitive overhead that users would take notes with pen and paper to help keep track of what they were doing~\cite{10.1145/800045.801580}.
The research literature featured a multitude of task-switching interface designs in the subsequent years, including Information Visualizer~\cite{robertson1993information}, Data  Mountain~\cite{robertson1998data}, and others (see~\cite{card1996webbook,czerwinski1999thumbnail,robertson2000taskgallery}).
Consumers would see the first GUI task-switching features as early as Windows 1.0 with Alt+Tab, although visual indicators wouldn't appear until Windows 3.0 in 1990~\cite{chen2009alttab} and the now-familiar thumbnail previews debuted later with Windows Vista in 2007~\cite{chen2009alttab}.
These task management interfaces support knowledge workers, who frequently switch between eight or more application windows~\cite{hutchings2004display} potentially hundreds of times every hour~\cite{olivier2006swish, leijten2014writing} to complete their tasks.





There is a growing body of research on using XR for knowledge work. 
Interfaces designed for this use case commonly focus on how the experience of working with desktop applications can be improved through using an XR headset, e.g., by providing a larger workspace for virtual monitors~\cite{daeijavad2024designing, queiroz2024xrknowledgework}.
This provides the same kind of task switching experience that we have come to expect through window-based applications on desktops.
However, there are many beneficial uses for knowledge work in XR beyond window-based applications~\cite{queiroz2024xrknowledgework}, so there is a need for more research on task switching interfaces that consider the unique affordances (and challenges) of XR. 
For instance, recent taxonomies of collaborative XR tasks have identified systematic frameworks for understanding how people can collaborate and interact in XR environments while promoting efficient communication and productivity~\cite{ghamandi2023and, ghamandi2024unlocking}. Additionally, some research in XR can be applied to serve these needs, though they do not often appear specifically in knowledge work settings.
Nam et al.~\cite{nam2019wedges} were perhaps the first to point out the limitations of the traditional XR ``one-world-at-a-time'' paradigm and presented the Worlds-in-Wedges technique whereby users could experience multiple views of data in different slices of their field of regard.
Das et al. explored finger-based techniques~\cite{das2024fingerworn} for quickly switching between augmented and virtual versions of an environment.
In Schjerlund et al.'s OVRLap, we find a technique for combining multiple immersive viewpoints through semi-transparent overlays, inspired by having multiple windows open in a desktop context~\cite{schjerlund2022ovrlap}.
Following that work and explicitly studying presence in world-switching contexts, Ablett et al. created portals with overlay techniques to preview other worlds and better enable multi-world presence~\cite{ablettSimultaneousPresenceContinuum2025}.
These techniques have contributed initial explorations into enabling task switching across complex XR environments.
With this paper, we hope to inspire and guide future work on transitional interfaces designed specifically to support the productive future of XR knowledge work.

\section{Conceptual Framework for Task Switching in XR Knowledge Work}

Our preliminary conceptual framework for XR task switching is illustrated in~\figref{fig:teaser}.
It consists of examples of task types commonly performed in XR knowledge work, factors affecting the user's performance when switching tasks, contextual factors relating to the user and the world they are in, and factors relating to transitioning between these contexts.
This framework provides a general structure for all of these components and can serve as a starting point for understanding the key factors involved in creating interfaces for supporting efficient and effective transitions between different XR tasks.
We then analyze how the principal characteristics of XR contexts, tasks, and transitions require interface explorations that extend beyond the capabilities of existing solutions for window and task management in desktop systems.

We use this framework to help generate four speculative scenarios of knowledge workers completing workflows that involve transitioning between different task types and contexts.
We further analyze our scenarios to identify the key features that need to be integrated into an XR operating system to enable efficient context switching.
We ultimately argue that the affordances and complexities of XR warrant further research into the task transition interfaces that will reify our vision of XR maturing into a daily productivity medium.

\subsection{Unique Challenges of XR Computing Contexts}

\newcommand{\tightcolorbox}[2]{%
  {\setlength{\fboxsep}{1.25pt}\colorbox{#1}{\textbf{\textsc{#2}}}}%
}

\newcommand{\cul}[3][red]{%
  \tikz[baseline=(X.base)]{
    \node[inner sep=0pt, outer sep=0.5pt] (X) {#2};
    \draw[#1,very thick] (X.south west) -- (X.south east);
  }%
}

\newcommand{\workContext}{\tightcolorbox{bgCtx}{Work Context}}
\newcommand{\taskFactors}{\tightcolorbox{bgTask}{Task Factors}}
\newcommand{\transitionFactors}{\tightcolorbox{bgTransition}{Transition Factors}}

\newcommand{\chair}{{\includegraphics[height=1em]{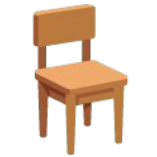}}}
\newcommand{\pin}{\includegraphics[height=1em]{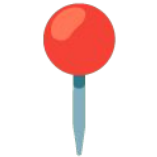}}
\newcommand{\social}{\includegraphics[height=0.9em]{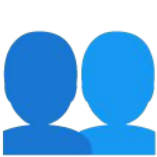}}
\newcommand{\self}{\includegraphics[height=1em]{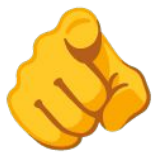}}
\newcommand{\brain}{\includegraphics[height=1em]{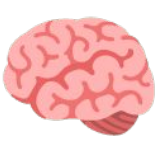}}
\newcommand{\physioI}{\includegraphics[height=0.9em]{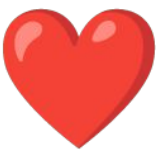}}
\newcommand{\analyzeI}{\includegraphics[height=1em]{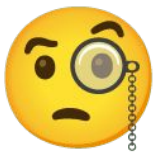}}
\newcommand{\infoRetrieveI}{\includegraphics[height=0.9em]{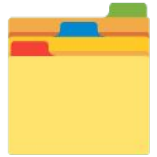}}
\newcommand{\authorI}{\includegraphics[height=1em]{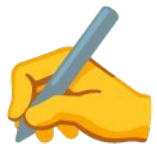}}
\newcommand{\evaluateI}{\includegraphics[height=0.9em]{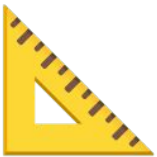}}
\newcommand{\learnI}{\includegraphics[height=1em]{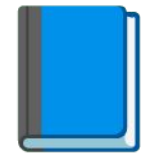}}
\newcommand{\collaborateI}{\includegraphics[height=1em]{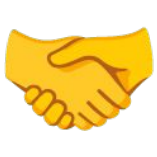}}
\newcommand{\asynccommI}{\includegraphics[height=0.8em]{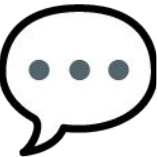}}
\newcommand{\planI}{\includegraphics[height=0.85em]{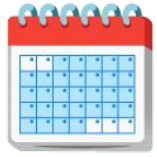}}
\newcommand{\similarityI}{\includegraphics[height=1em]{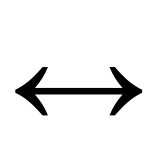}}
\newcommand{\complexityI}{\includegraphics[height=0.95em]{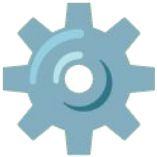}}
\newcommand{\recoveryI}{\includegraphics[height=0.85em]{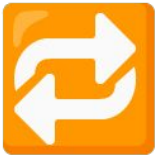}}
\newcommand{\tasktimeI}{\includegraphics[height=0.9em]{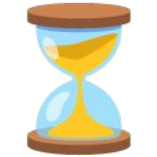}}
\newcommand{\transitionInitiatorI}{\includegraphics[height=0.9em]{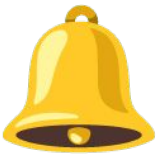}}
\newcommand{\transitionEffectI}{\includegraphics[height=0.9em]{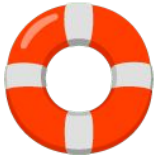}}

\newcommand{\iconlabel}[5]{\hspace{#3}\raisebox{#5}{#1}\hspace{#4}{\small\textbf{#2}}}

\newcommand{\cognitivestate}{\iconlabel{\brain}{Cognitive State}{-0.1em}{-0.0em}{-0.2em}}
\newcommand{\physiologicalstate}{\iconlabel{\physioI}{Physiological State}{-0.2em}{-0.0em}{-0.2em}}
\newcommand{\physicalstate}{\iconlabel{\chair}{Physical States}{-0.2em}{-0.1em}{-0.2em}}
\newcommand{\environment}{\iconlabel{\pin}{Environment}{-0.2em}{-0.1em}{-0.2em}}
\newcommand{\socialpresence}{\iconlabel{\social}{Social Presence}{0em}{0.1em}{-0.1em}}
\newcommand{\embodiment}{\iconlabel{\self}{Self Embodiment}{0em}{0.1em}{-0.2em}}
\newcommand{\analyze}{\iconlabel{\analyzeI}{Analyze}{-0.2em}{-0.0em}{-0.2em}}
\newcommand{\infoRetrieve}{\iconlabel{\infoRetrieveI}{Information Retrieval \& Organization}{-0.2em}{-0.0em}{-0.2em}}
\newcommand{\author}{\iconlabel{\authorI}{Author}{-0.2em}{-0.0em}{-0.2em}}
\newcommand{\evaluate}{\iconlabel{\evaluateI}{Evaluate}{-0.2em}{-0.0em}{-0.2em}}
\newcommand{\learn}{\iconlabel{\learnI}{Learn}{-0.2em}{-0.0em}{-0.2em}}
\newcommand{\collaborate}{\iconlabel{\collaborateI}{Meet \& Collaborate}{-0.2em}{-0.0em}{-0.2em}}
\newcommand{\asynccomm}{\iconlabel{\asynccommI}{Asynchronous Communication}{-0.2em}{-0.0em}{-0.1em}}
\newcommand{\plan}{\iconlabel{\planI}{Plan}{-0.2em}{-0.0em}{-0.1em}}
\newcommand{\similarity}{\iconlabel{\similarityI}{Similarity}{-0.2em}{-0.1em}{-0.2em}}
\newcommand{\complexity}{\iconlabel{\complexityI}{Complexity}{-0.1em}{-0.0em}{-0.2em}}
\newcommand{\recovery}{\iconlabel{\recoveryI}{Recovery}{-0.0em}{-0.0em}{-0.15em}}
\newcommand{\tasktime}{\iconlabel{\tasktimeI}{Time}{-0.1em}{-0.1em}{-0.15em}}
\newcommand{\transitionInitiator}{\iconlabel{\transitionInitiatorI}{Transition Initiator}{-0.1em}{-0.05em}{-0.1em}}
\newcommand{\transitionEffect}{\iconlabel{\transitionEffectI}{Transition Effect}{-0.1em}{-0.0em}{-0.2em}}

\renewcommand{\cognitivestate}{\iconlabel{\brain}{Cognitive State}{-0.1em}{-0.0em}{-0.2em}}
\renewcommand{\physiologicalstate}{\iconlabel{\physioI}{Physiological State}{-0.2em}{-0.0em}{-0.2em}}
\renewcommand{\physicalstate}{\iconlabel{\chair}{Physical State}{-0.2em}{-0.1em}{-0.2em}}
\renewcommand{\environment}{\iconlabel{\pin}{Environment}{-0.2em}{-0.1em}{-0.2em}}
\renewcommand{\socialpresence}{\iconlabel{\social}{Social Presence}{0em}{0.1em}{-0.1em}}
\renewcommand{\embodiment}{\iconlabel{\self}{Self Embodiment}{0em}{0.1em}{-0.2em}}
\renewcommand{\analyze}{\iconlabel{\analyzeI}{Analyze}{-0.2em}{-0.0em}{-0.2em}}
\renewcommand{\infoRetrieve}{\iconlabel{\infoRetrieveI}{Information Retrieval \& Organization}{-0.2em}{-0.0em}{-0.2em}}
\renewcommand{\author}{\iconlabel{\authorI}{Author}{-0.2em}{-0.0em}{-0.2em}}
\renewcommand{\evaluate}{\iconlabel{\evaluateI}{Evaluate}{-0.2em}{-0.0em}{-0.2em}}
\renewcommand{\learn}{\iconlabel{\learnI}{Learn}{-0.2em}{-0.0em}{-0.2em}}
\renewcommand{\collaborate}{\iconlabel{\collaborateI}{Meet \& Collaborate}{-0.2em}{-0.0em}{-0.2em}}
\renewcommand{\asynccomm}{\iconlabel{\asynccommI}{Asynchronous Communication}{-0.2em}{-0.0em}{-0.1em}}
\renewcommand{\plan}{\iconlabel{\planI}{Plan}{-0.2em}{-0.0em}{-0.1em}}
\renewcommand{\similarity}{\iconlabel{\similarityI}{\cul[bgTask]{Similarity}}{-0.2em}{-0.1em}{-0.2em}}
\renewcommand{\complexity}{\iconlabel{\complexityI}{\cul[bgTask]{Complexity}}{-0.1em}{-0.0em}{-0.2em}}
\renewcommand{\recovery}{\iconlabel{\recoveryI}{\cul[bgTask]{Recovery}}{-0.0em}{-0.0em}{-0.15em}}
\renewcommand{\tasktime}{\iconlabel{\tasktimeI}{\cul[bgTask]{Time}}{-0.1em}{-0.1em}{-0.15em}}
\renewcommand{\transitionInitiator}{\iconlabel{\transitionInitiatorI}{\cul[bgTransition]{Transition Initiator}}{-0.1em}{-0.05em}{-0.1em}}
\renewcommand{\transitionEffect}{\iconlabel{\transitionEffectI}{\cul[bgTransition]{Transition Effect}}{-0.1em}{-0.0em}{-0.2em}}

\renewcommand{\cognitivestate}{\cul[bgCtx]{\iconlabel{\brain}{Cognitive State}{-0.1em}{-0.0em}{-0.2em}}}
\renewcommand{\physiologicalstate}{\cul[bgCtx]{\iconlabel{\physioI}{Physiological State}{-0.2em}{-0.0em}{-0.2em}}}
\renewcommand{\physicalstate}{\cul[bgCtx]{\iconlabel{\chair}{Physical State}{-0.2em}{-0.1em}{-0.2em}}}
\renewcommand{\environment}{\cul[bgCtx]{\iconlabel{\pin}{Environment}{-0.2em}{-0.1em}{-0.2em}}}
\renewcommand{\socialpresence}{\cul[bgCtx]{\iconlabel{\social}{Social Presence}{0em}{0.1em}{-0.1em}}}
\renewcommand{\embodiment}{\cul[bgCtx]{\iconlabel{\self}{Self Embodiment}{0em}{0.1em}{-0.2em}}}
\renewcommand{\analyze}{\cul[bgTask]{\iconlabel{\analyzeI}{Analyze}{-0.2em}{-0.0em}{-0.2em}}}
\renewcommand{\infoRetrieve}{\cul[bgTask]{\iconlabel{\infoRetrieveI}{Information Retrieval \& Organization}{-0.2em}{-0.0em}{-0.2em}}}
\renewcommand{\author}{\cul[bgTask]{\iconlabel{\authorI}{Author}{-0.2em}{-0.0em}{-0.2em}}}
\renewcommand{\evaluate}{\cul[bgTask]{\iconlabel{\evaluateI}{Evaluate}{-0.2em}{-0.0em}{-0.2em}}}
\renewcommand{\learn}{\cul[bgTask]{\iconlabel{\learnI}{Learn}{-0.2em}{-0.0em}{-0.2em}}}
\renewcommand{\collaborate}{\cul[bgTask]{\iconlabel{\collaborateI}{Meet \& Collaborate}{-0.2em}{-0.0em}{-0.2em}}}
\renewcommand{\asynccomm}{\cul[bgTask]{\iconlabel{\asynccommI}{Asynchronous Communication}{-0.2em}{-0.0em}{-0.1em}}}
\renewcommand{\plan}{\cul[bgTask]{\iconlabel{\planI}{Plan}{-0.2em}{-0.0em}{-0.1em}}}
\renewcommand{\similarity}{\cul[bgTask]{\iconlabel{\similarityI}{Similarity}{-0.2em}{-0.1em}{-0.2em}}}
\renewcommand{\complexity}{\cul[bgTask]{\iconlabel{\complexityI}{Complexity}{-0.1em}{-0.0em}{-0.2em}}}
\renewcommand{\recovery}{\cul[bgTask]{\iconlabel{\recoveryI}{Recovery}{-0.0em}{-0.0em}{-0.15em}}}
\renewcommand{\tasktime}{\cul[bgTask]{\iconlabel{\tasktimeI}{Time}{-0.1em}{-0.1em}{-0.15em}}}
\renewcommand{\transitionInitiator}{\cul[bgTransition]{\iconlabel{\transitionInitiatorI}{Transition Initiator}{-0.1em}{-0.05em}{-0.1em}}}
\renewcommand{\transitionEffect}{\cul[bgTransition]{\iconlabel{\transitionEffectI}{Transition Effect}{-0.1em}{-0.0em}{-0.2em}}}

As visualized in the conceptual framework illustrated in~\figref{fig:teaser}, knowledge workers' experiences switching between tasks can be described in terms of the \workContext{} in which they are completing a task, various associated \taskFactors{}, and \transitionFactors{} having to do with how they move between tasks.
Below, we describe examples of the framework factors and provide examples of how they are differently complex in XR compared to non-XR productivity settings.

A user's \workContext{} includes various \textit{User Factors} that involve viewing and interacting with digital content from a different \cognitivestate{}, \physiologicalstate{}, and/or \physicalstate{}.
For \physicalstate{}, users may switch between different postures.
For example,
desktop-based knowledge workers might utilize a desk with an adjustable height to switch between sitting and standing, or they might jot down notes or draft an email as they walk.
In XR, there are fewer physical constraints for where content can be displayed, which can lead to content being placed anywhere around the user and in different reference frames.

XR application \textit{World Factors} also involve much more complex display and interaction compared to desktop applications.
In desktop settings, application windows are bounded and stabilized by the monitor's display space.
However, in XR, a user may experience substantial differences in terms of \environment{}, \embodiment{}, and \socialpresence{} when switching between tasks.
For example, switching from one virtual world to another (and back again) can introduce substantial changes in both content and format --- users can find themselves in entirely different surroundings with virtual objects that can be positioned in all directions.
Additionally, the XR user's \embodiment{} is more dynamic.
They may prefer to embody different avatars in different worlds, e.g., a realistic avatar when they meet with their boss and a dragon when they take a break to play a game with friends.
Their interaction techniques may also change so that they may best accomplish different tasks. For example, prior work has shown that interaction techniques should be adapted to different task contexts and user preferences to maintain efficiency~\cite{maslych2024research, hmaiti2024visual}.
On the contrary, the desktop user's \embodiment{} is generally very stable across applications --- the user remains themselves, and their method of interaction with application content is (very often) limited to keyboard and mouse.

Last, XR affords richer social information about users, which creates the perception of high \socialpresence{}, the experience of being in the same room as someone compared to interacting with them through a screen.
In desktop video-conferencing meetings, looking away from the camera to look at a different app window is not very visible or disruptive because users do not always make direct eye contact with their webcam, all other applications occupy the display space generally limited to the forward direction of the desktop user, and the user's interactions are often hidden from view (i.e., collaborators do not see the user's mouse when they are not sharing their screen, and the user's hands are often not in frame).
However, when someone multitasks in a VR meeting like in Scenario D, it will become obvious to the other meeting attendees that the multitasker is not always looking at the person speaking and their avatar may even appear to behave strangely as they interact with other worlds.
In this case, it becomes necessary to redirect the user's avatar whenever they choose to multitask so they do not disrupt the meeting.


Various \transitionFactors{} affect the user's experience when switching tasks.
These include the \transitionInitiator{} as well as the 
\transitionEffect{} itself.
The \transitionInitiator{} can be internal, e.g., the user themselves chooses to switch tasks because they have finished one task or they have another task scheduled.
It may also be external, as in the case of an interruption from a colleague.
Using XR for knowledge work in real office environments adds a complicating dimension to the management of interruptions: handling cross-reality interruptions~\cite{gottsacker2022cues}.
When someone is immersed in an XR environment, they may not be able to see a real-world interrupter due to virtual content occluding them.
Researchers have explored this problem considerably, 
and modern consumer headsets incorporate bystander-sensing technologies.
There may be an issue from the other side of this interaction, however: the real-world interrupter may have trouble determining when it is a good time to interrupt an XR user because they cannot observe what the XR user is doing or fully observe their level of engagement~\cite{gottsacker2022cues, gottsacker2025examining}.

The \transitionEffect{} can be more complex in XR compared to desktop environments.
When a desktop user triggers a task switch, the visual effect of the transition is limited to activating a visual indicator on the window that is now in focus (e.g., a drop-shadow or colored outline) or a swiping animation that replaces the monitor's contents with a different workspace.
As outlined in the description of the \workContext{}, the content that an XR user interacts with for a given task is more complex --- switching tasks potentially involves switching entire worlds.
Different kinds of transitions between worlds can have different effects on the user's experience~\cite{feld2024simple, GottsackerResidue}.

\taskFactors{} also have an impact on the user's task switching experience.
For example, when a user is working on a task under \tasktime{} pressure (i.e., a deadline) and switches to a new task without finishing the first task, they will perform more poorly on the second task~\cite{leroy_tasks_2018}.
Some other factors that have been shown to have an effect on performance include task \complexity{}, the \similarity{} between tasks before and after a transition, and the \tasktime{} spent on tasks~\cite{trafton2007taskinterruptions}.
Particularly of interest to us is when a user needs assistance with the \recovery{} of some context in order to complete their task, as in the case of them resuming a task and recalling the trajectory of their work, or coming up to speed on changes that a collaborator made to a project while they were away.
Research has demonstrated the benefits of cues to assist with resuming simple spatial tasks~\cite{bahnsen2024arresumption}.

\subsection{Representative XR Knowledge-Work Scenarios}
\label{sec:scenarios}


We describe four speculative scenarios that illustrate frequent occurrences of task switching in future XR workflows based on day-to-day tasks of contemporary knowledge work. 
To create these scenarios, we first consulted Reinhardt et al.'s typology of knowledge worker tasks~\cite{reinhardt2011knowledge}.
We focused on task types for which researchers have demonstrated unique benefits of XR (for examples of promising XR knowledge work applications, see~\cite{queiroz2024xrknowledgework}).
Additionally, we left out some of the task types outlined in Reinhardt et al.'s typology~\cite{reinhardt2011knowledge} that do not involve substantive spatial components and for which we do not have ready examples of XR contributing promising transformational power (e.g., Expert Search, Service Search).
Thus, we arrived at the following task types (also shown in~\figref{fig:teaser}) from which we derived our scenarios: \analyze{}, \infoRetrieve{}, \author{}, \evaluate{}, \learn{}, \collaborate{}, \asynccomm{}, and \plan{}.


\paragraph{\textbf{A. 3D Scene Development Workflow}}
An architect designs a building and a custom VR tour for a client.
The tour includes viewing the building from a bird's-eye view with nearby buildings visible, from the street level in front of the building, and inside the front room.
The architect switches between all of these views, walking around the full-scale views and adjusting positions and rotations of objects.
She notices that the hardwood floor texture has some visual artifacts in it, so she selects the texture and edits the base color to try to make the issues less visible.
She is still not satisfied with how it looks, so she opens up a photo editing application window to use a blurring brush to ``smooth out'' the image.
She returns to her desk, sits down, and edits the texture using her mouse, as she prefers a stabilized, elbow-supported posture for precise cursor control.

\paragraph{\textbf{B. Learning from Immersive Data Story}}

A business analyst reviews an immersive data story featuring complex 3D market trend visualizations, while simultaneously drafting a report. During a critical segment requiring deep analysis and concentration, his manager interrupts via phone call to request scheduling meetings of performance reviews. The analyst exits the immersive environment, completes the scheduling task, and attempts to resume the data story. However, he struggles to remember how the complicated parts of the story were coming together. Here, AI-generated in-situ resumption cues (e.g., keyframe-based spatial visualizations) restore both his cognitive and spatial context, enabling efficient task continuation.

\paragraph{\textbf{C. Collaborative Brainstorming for Product Design}}
Two fashion designers are having a remote brainstorming session in VR.
They use a virtual whiteboard to first sketch out some ideas for a new dress.
One designer uses a generative AI to create 3D models guided by their sketches and descriptions.
He manipulates a model in his hands, looking at it from different angles and modifying its shape.
He wants to add a small button to one part of the garment, so he activates a manipulation tool implemented in a 3D modeling application to assist with fine-grained positioning and rotation.
The other designer is reminded of an interesting car that she saw on vacation several years ago and wants to bring it up as inspiration.
She transitions to her private memory palace where she has cataloged some of her favorite life experiences through images and scans.
She navigates though some virtual corridors while keeping an audio channel open to her collaborator.
While she is ``away'' from the meeting room, her avatar loops through an idle animation and becomes translucent to indicate her decreased level of presence.
Upon returning to the brainstorming room, she is re-oriented to the direction that her avatar was facing before she left.

\paragraph{\textbf{D. Multitasking During a Meeting}}
A professor attends a virtual department meeting in VR where she's not expected to actively participate. She activates ``meeting-in-miniature" mode, transitioning to AR where the meeting table and attendee avatars shrink to desktop size while maintaining full audio. In the meeting environment, her avatar displays ambient presence through idle animations that track only essential social cues like nodding, allowing her to multitask on email and scheduling without disrupting the meeting's social dynamics.
When the department head announces significant budget cuts, she taps the miniaturized meeting to return to full-scale VR engagement. Upon hearing about canceled faculty searches, she immediately initiates a private side-meeting with a colleague who had extended an offer to a top candidate. The system seamlessly creates a separate meeting space while maintaining avatar copies in the original meeting, with the department meeting audio continuing as ambient background sound.
After their brief private discussion, both return to email tasks while the meeting continues in miniature mode. When the department head shares a budget document, a mail icon appears over the miniaturized meeting space, which the professor taps to open the document on her desktop screen.

\subsection{Key Features to Achieve Fluid XR Task Switching}
\label{sec:keyfeatures}

\textbf{The 3D scene development workflow (Scenario A)} exemplifies the need for transitions that preserve both geometric and interaction contexts across different work modes. When the architect transitions from immersive 3D manipulation to precision 2D texture editing, the system must maintain spatial awareness of the selected object, while simultaneously adapting \embodiment{} (e.g., tracked hand to precise cursor) and display characteristics (immersive to windowed). This necessitates interface mechanisms that can dynamically bridge spatial and traditional computing paradigms without losing track of task context and tracked \physicalstate{}. Then, the transition back to immersive viewing must restore not only the spatial arrangement but also the user's previous viewpoint and interaction state, which in itself suggests the need for a persistent spatial bookmarking and context aware resumption cues. 

\textbf{The learning from immersive data story (Scenario B)} highlights the complexity of managing an external \transitionInitiator{} in immersive environments. When the analyst's manager interrupts the data story viewing, the system must gracefully handle the transition from focused individual learning to collaborative task execution, then back to the original learning context. This requires sophisticated interruption management interfaces that can: (1) capture the current immersive \cognitivestate{} and \physicalstate{} including temporal position, visual focus, and cognitive state of the user; (2) provide rapid context switching to accommodate urgent requests; and then (3) offer intelligent \recovery{} for task resumption support appropriate for \complexity{} and \similarity{} of tasks. Additionally, the interface must balance notification awareness and preserving immersion, potentially through ambient peripheral displays or haptic cues that maintain immersion, while signaling external demands.

\textbf{The collaborative brainstorming for product design (Scenario C)} shows the need for \socialpresence{} management interfaces that can handle multiple concurrent collaborative contexts. When one designer transitions from shared whiteboard to 3D model manipulation, the system must seamlessly bridge between two dissimilar tools, the 2D sketching tools and spatial 3D modeling interfaces while maintaining collaborative awareness. The subsequent transition to the private memory palace requires even more complex interface mechanisms: preserving audio communication channels, managing avatar presence states (e.g., translucent materials, idle animations), and maintaining the shared workspace state during the designer's absence. The interface must support contextual presence indicators that communicate availability and engagement levels to collaborators while enabling private exploration. Upon return, the system must provide spatial re-orientation that restores both the user's physical positioning and their \cognitivestate{} to the collaborative flow. This requires interfaces that can manage multiple simultaneous spatial contexts: the shared brainstorming \environment{}, the 3D modeling space, and the private memory palace, while preserving social presence and collaboration continuity.

\textbf{The multi-tasking during a meeting (Scenario D)} shows the need for scale-adaptive interfaces that can dynamically adapt to different engagement levels, while maintaining \socialpresence{} in concurrent tasks. The transition from full-scale VR meeting to AR meeting-in-miniature mode requires interfaces that can seamlessly connect immersive and ambient engagement paradigms. The system must maintain audio perspective, while spatially compressing the meeting \environment{} and managing avatar presence across different scales. More complex still is the creation of parallel social contexts when the professor initiates a private meeting with a colleague; the interface must support social context forking, where avatar copies maintain presence in the original meeting while the actual users engage in a separate conversation. 
XR interfaces must provide clear affordances for managing these parallel social streams, while making sure that users remain aware of their various presence commitments and are able to manage their distributed attention across multiple collaborative contexts.

\section{XR Productivity Design Challenges}



In the representative scenarios we described (\secref{sec:scenarios}) and key features to achieve them (\secref{sec:keyfeatures}), we highlight several interface mechanisms and transition types that are currently missing from most XR systems. These gaps exist partly because XR is still relatively new in the context of knowledge work. However, progress is also hindered by attempts to replicate desktop functionality --- where applications are separated into windows --- rather than embracing the interaction paradigms, spatial contexts, and collaborative states that XR affords. Below, we outline key challenges that offer opportunities for XR-first design solutions.


\subsection{Cross-Application Interoperability}
Desktop productivity relies heavily on seamless content transfer through standardized copy-paste mechanisms enabled by universal keyboard shortcuts. XR systems lack equivalent functionality, forcing users to manually recreate or export content when switching between applications. This limitation becomes particularly problematic when transitioning between spatial and 2D content representations, where traditional clipboard metaphors are inadequate.

\noindent\textbf{Design Opportunity}: XR systems require new paradigms for content transfer that leverage spatial interaction. This could include gesture-based content selection, spatial and chronological ``clipboard" areas where users can temporarily store 3D objects, spatial arrangements or other relevant information, and cross-application content bridges that maintain semantic and contextual meaning across different representational formats. 

\subsection{Cross-Device Interoperability}
To complete daily tasks, knowledge workers often switch between stationary and mobile devices, as well as input and output hardware designed for specific purposes. Current XR systems are limited in interoperability, locking users in tight ecosystems with limited hardware options and communication channels. This forces users to find cumbersome workarounds for content transfer and state synchronization across XR devices.

\noindent\textbf{Design Opportunity}: XR devices should support transfer, playback and interactions with content from other computing mediums, leveraging XR affordances for work that other devices fail to support seamlessly. For example, XR should allow an architect to easily add spatial annotations to a building plan while receiving feedback from clients attending virtually from desktop and mobile. Later, viewing annotations made in XR through any medium would not result in lost context, which could have occurred if annotations were made in 2D.

\subsection{Persistent Application Access}
The absence of taskbar-equivalent interfaces in XR creates an important friction for application switching. XR users must navigate through multiple menu layers or exit curent applications entirely to access alternative tools, disrupting workflow continuity and increasing cognitive load.

\noindent\textbf{Design Opportunity}: XR environments should provide customizable, persistent access to active and frequently used applications through spatial interfaces that act as personal ``inventories" of tools and workspaces. These could be implemented as user-configurable spatial panels, gesture-accessible application launchers, or context-aware tool recommendations that adapt to currently performed work modes.

\subsection{Efficient Command Invocation}
Desktop keyboard shortcuts enable rapid command execution across applications. XR's reliance on hand controllers and gesture recognition creates challenges for implementing equivalent shortcut systems, particularly given the need for gestures that are both memorable and distinguishable across different rich contexts that XR affords.

\noindent\textbf{Design Opportunity}: XR systems require development of standardized gesture vocabularies and voice command protocols that can function consistently across applications. This includes investigating multi-modal shortcut combinations (e.g., gesture + voice, gaze + gesture, etc.) and adaptive command interfaces that adjust to user preferences and capabilities.

\subsection{Context-Aware Task Resumption}
Effective task resumption requires presenting sufficient contextual information to help users quickly re-engage with interrupted work. XR's spatial nature provides both opportunities and challenges for context preservation, as traditional desktop indicators (window titles and previews, taskbar icons) may not adequately convey the state of immersive environments.

\noindent\textbf{Design Opportunity}: XR systems should leverage spatial memory and embodied cues to support task resumption. This includes developing adaptive information presentation formats that adjust to different task factors, including task \tasktime{} considerations (e.g., duration, deadlines, and time away from the task), as well as personalization mechanisms that accommodate individual differences in spatial cognition and preference.

\subsection{Embodiment and Tool Coherence}
User embodiment in XR must reflect available interaction capabilities and current work context. Inconsistent or inappropriate embodiment can create confusion about available actions and tools, particularly during task transitions that require different interaction modalities and techniques.

\noindent\textbf{Design Opportunity}: XR systems should implement dynamic embodiment that adapts to work context, clearly indicating available tools and interaction modes through avatar representation and environmental cues. This includes developing transition animations and preview systems that prepare users for embodiment changes across different work modes and inform them about the available tools. Furthermore, the design of these embodiment transitions must accommodate individual preferences and task requirements. Users may prefer a rapid, minimal \transitionEffect{} for efficiency in some cases, and gradual, sensory- and information-rich transitions in other cases, accounting for \cognitivestate{} and \physiologicalstate{}. 



\subsection{XR System Design Implications}
The design challenges we outline are ambitious, and current technical limitations mean they cannot be fully realized yet. Nonetheless, these constraints present valuable avenues for research exploration, guiding future investigations toward overcoming these barriers. These challenges point toward the need for XR operating systems and validated interaction frameworks that support productivity workflows at a fundamental level. Rather than expecting individual applications to solve these problems in isolation, system-level solutions can provide consistent, learnable interaction paradigms that work across different XR productivity tools. This includes developing standardized APIs for content transfer, persistent interface elements, and cross-application state management that can enable the fluid, efficient task switching that knowledge workers require.

\noindent\textbf{Design Opportunity}: 
The productivity workflows we identified demand system-level support for concurrent application execution, where the OS manages decoupling of interaction techniques, locomotion methods, and user representation at runtime rather than expecting individual applications to solve these problems in isolation. Moreover, future XR systems need architectural approaches that decouple core interaction components, including interaction techniques, locomotion systems, and avatar representation, from specific applications. This would enable users to maintain consistent interaction preferences across different work contexts while allowing applications to focus on domain-specific functionality rather than reinventing basic common interaction paradigms. 

\section{Conclusion}

The future of productive XR lies not in replicating the past, but in embracing the transformative potential of immersive computing. 
\textbf{The success of XR for knowledge work will be measured not by how closely XR approximates desktop productivity, but by how effectively it enables new forms of collaborative, spatial, and embodied work that are impossible in traditional computing environments.}
Our preliminary conceptual framework and analysis of representative productivity scenarios in XR reveal specific gaps that prevent XR from supporting the fluid task switching knowledge workers need. We identified several critical design challenges, from cross-application interoperability to context-aware resumption, which demand novel interaction techniques and system-level solutions rather than application-specific workarounds. 

While our framework extends existing thinking on knowledge work in XR, it remains at the conceptual stage and has not yet been empirically validated. 
Addressing this shortcoming will require systematic empirical studies, including controlled experiments, field studies, and longitudinal deployments, to test the framework's assumptions, refine its components, and measure its effectiveness in diverse XR knowledge work setups.
Only through coordinated effort to move beyond ad-hoc prototypes toward validated and standardized interaction paradigms can we mature XR into a medium truly suitable for sustained knowledge work. Realizing this vision will require XR operating systems that decouple interaction techniques, locomotion methods, and user representation from individual applications, enabling consistent workflows across diverse productivity tools.

\acknowledgments{
This material includes work supported in part by the National Science Foundation under Award Number 2235066 (Dr. Han-Wei Shen, IIS); the Office of Naval Research under Award Numbers N000142512159 and N000142512245 (Dr. Peter Squire, Code 34); and the AdventHealth Endowed Chair in Healthcare Simulation (Prof. Welch).
}

\bibliographystyle{abbrv-doi}

\bibliography{template}


\end{document}